\documentclass[aps,prd,nofootinbib,showpacs,superscriptaddress,floatfix, preprint]{revtex4-1}
\pdfoutput=1
\usepackage[utf8]{inputenc}
\usepackage{amsmath,amssymb}
\usepackage{epsfig}
\usepackage{bbold}
\usepackage{graphicx}
\usepackage[usenames,dvipsnames]{color}
\usepackage[colorlinks,citecolor=blue]{hyperref}
\usepackage{color}
\usepackage{subfigure}
\usepackage{hhline}
\usepackage{multirow}
\usepackage{subfigure}
\begin{document}

	\title{Survey of Texture Zeros with Light Dirac Neutrinos}
	
	%%%%%%%%%   Authors   %%%%%%%%%%%%
	\author{Happy Borgohain}
	\email{happy@iitg.ac.in}
%	\affiliation{Department of Physics, Indian Institute of Technology Guwahati, Assam 781039, India}

	\author{Debasish Borah}
	\email{dborah@iitg.ac.in}
	\affiliation{Department of Physics, Indian Institute of Technology Guwahati, Assam 781039, India}
	
\begin{abstract}
We classify all possible texture zeros in light neutrino mass matrix in diagonal charged lepton basis by considering Dirac nature of light neutrinos. For Hermitian nature of neutrino mass matrix, the number of possible texture zeros remain same as Majorana texture zeros, but with less free parameters due to the absence of additional Majorana CP phases. Relaxing the Hermitian nature of neutrino mass matrix lead to many more possibilities and freedom due to additional CP phases, mixing angles not constrained by neutrino data. While we find that none of the texture zeros in Hermitian case are allowed, in non-Hermitian case some textures even with four zeros are allowed. While most of the one-zero, two-zero and three-zero textures in this case are allowed, four-zero textures are tightly constrained with only 6 allowed out of 126 possibilities. The allowed textures also give interesting correlations between light neutrino parameters with sharp distinctions between normal and inverted mass ordering. Many of these allowed textures also saturate the Planck 2018 upper bound on the sum of absolute neutrino masses.
\end{abstract}
	\maketitle

\section{Introduction}
In spite of significant advancement in understanding the physics of light neutrinos over last several decades, there are still many unanswered questions related to them. While experimental evidences gathered over a long period of time have confirmed that light neutrinos have non-zero but tiny masses and large mixing \cite{Mohapatra:2005wg, Tanabashi:2018oca, deSalas:2017kay, Esteban:2018azc}, yet we do not know their fundamental origin. Due to the absence of right handed neutrinos in the standard model (SM) of particle physics, light neutrinos remain massless due to absence of any renormalisable coupling with the Higgs field thereby resulting in vanishing mixing as well. In addition to this, there are still uncertainties in the octant of atmospheric mixing angle. Also, since only two mass squared differences are experimentally measured, we do not have any idea about the lightest neutrino mass leading to the possibility of both normal ordering (NO): $m_1 < m_2 < m_3$ as well as inverted ordering: $m_3 < m_1 < m_2$ (see table \ref{tab:data1} for latest global fit data). On the other hand, tritium beta decay experiment KATRIN has recently measured an upper limit of 1.1 eV on absolute neutrino mass scale \cite{Aker:2019uuj} which is however, weaker than the upper limits from cosmology. Latest data from Planck collaboration constrains the sum of absolute neutrino masses as $\sum_i \lvert m_i \rvert < 0.12$ eV \cite{Aghanim:2018eyx}. In spite of recent advances in determining the leptonic Dirac CP phase \cite{Abe:2019vii} suggesting a maximal CP violation, it is not yet measured with sufficient accuracy to be considered as a discovery.

In addition to these, the nature of light neutrinos: Majorana or Dirac also remain undetermined at neutrino oscillation experiments. While positive signal at neutrinoless double beta decay ($0\nu \beta \beta$) experiments can confirm the Majorana nature of light neutrinos, we do not yet have any positive signal with latest data from KamLAND-Zen experiment \cite{KamLAND-Zen:2016pfg} providing the most stringent lower bound on the half-life of $0 \nu \beta \beta$ using $^{136} Xe$ nucleus as $ \rm T_{1/2}^{0\nu}>1.07\times 10^{26}$ year at $ 90\%$ CL. While the absence of positive signal at $0\nu \beta \beta$ experiments do not necessarily rule out the Majorana nature of light neutrinos, yet it is very motivating to consider the possibility of light Dirac neutrinos as well. While in SM, incorporating right handed neutrinos do lead to a Dirac Yukawa coupling with Higgs, the gauge symmetry of the model does not prevent a Majorana mass term for singlet right handed neutrinos thereby leading to Majorana light neutrinos. Even if the Majorana mass term is assumed to be absent, the Dirac Yukawa couplings are required to be tuned at the level of $< 10^{-12}$ in order to generate sub-eV light neutrino masses. Several beyond standard model (BSM) have been proposed in order to explain light Dirac neutrino masses in the last several years \cite{Babu:1988yq, Peltoniemi:1992ss, Chulia:2016ngi, Aranda:2013gga, Chen:2015jta, Ma:2015mjd, Reig:2016ewy, Wang:2016lve, Wang:2017mcy, Wang:2006jy, Gabriel:2006ns, Davidson:2009ha, Davidson:2010sf, Bonilla:2016zef, Farzan:2012sa, Bonilla:2016diq, Ma:2016mwh, Ma:2017kgb, Borah:2016lrl, Borah:2016zbd, Borah:2016hqn, Borah:2017leo, CentellesChulia:2017koy, Bonilla:2017ekt, Memenga:2013vc, Borah:2017dmk, CentellesChulia:2018gwr, CentellesChulia:2018bkz, Han:2018zcn, Borah:2018gjk, Borah:2018nvu, CentellesChulia:2019xky,Jana:2019mgj, Borah:2019bdi, Dasgupta:2019rmf, Correia:2019vbn, Ma:2019byo, Ma:2019iwj, Baek:2019wdn, Saad:2019bqf, Jana:2019mez, Nanda:2019nqy, Calle:2019mxn, Gu:2019yvw, Han:2020oet, Borah:2020boy}. In most of these models, Majorana mass of right handed neutrinos are forbidden due to additional symmetries and some of them also explains the natural origin of tiny Dirac neutrino masses via some kind of seesaw mechanism.

Motivated by the growing interests in light Dirac neutrinos and associated phenomenology, we intend to study the possible structures of light Dirac neutrino mass matrix in a model independent manner. In particular, we focus on the possibility of texture zeros in light Dirac neutrino mass matrix. While the light neutrino mass matrix can have many more free parameters compared to the experimentally determined five neutrino parameters, possibility of texture zeros, based on some underlying symmetries, can reduce the number of such free parameters. In fact, if the texture zero conditions are predictive enough, it is also possible to either rule them out based on latest neutrino data or obtain some testable predictions for unknown neutrino parameters like CP phase, octant of atmospheric mixing angle, mass ordering etc. While most of the analysis related to texture zeros\footnote{For a comprehensive review, please refer to \cite{Ludl:2014axa}.} have focused on Majorana nature of light neutrinos \cite{Xing:2002ta, Frampton:2002yf, Meloni:2014yea, Fritzsch:2011qv, Alcaide:2018vni, Zhou:2015qua, Bora:2016ygl, Singh:2016qcf, Ahuja:2017nrf, Borah:2015vra, Kalita:2015tda, Borgohain:2018lro, Barreiros:2018ndn, Barreiros:2018bju, Borgohain:2019pya, Borgohain:2020now, Ghosh:2012pw,Ghosh:2013nya,Zhang:2013mb,Nath:2015emg,Borah:2016xkc,Borah:2017azf, Sarma:2018bgf}, a relatively fewer number of attempts have been made to discuss the possibilities for light Dirac neutrinos \cite{Ahuja:2009jj, Liu:2012axa, Verma:2013cza, Fakay:2014nea, Sharma:2014wpa, Cebola:2015dwa, Ahuja:2016san, Singh:2018lao, Ahuja:2018fmw, Benavides:2020pjx}. Here we obtain an exhaustive list of allowed and disallowed texture zero classes for light Dirac neutrinos in the light of latest neutrino data. To begin with, we assume the light neutrino mass matrix to be Hermitian and find only a few one-zero textures to be allowed out of all possibilities. We then relax this assumption and consider the most general Dirac neutrino mass matrix which offers many more possibilities of texture zeros. We show that even a few three zero textures are also allowed in such a scenario. Apart from preparing such an exhaustive list of allowed textures, we also find interesting correlations and predictions for light neutrino parameters for some of the allowed textures.

This paper is organized as follows. In next section \ref{texture}, we discuss different possible textures for Hermitian case and then move onto the discussion of general structure of Dirac neutrino mass matrix in section \ref{sec3a}. In section \ref{result} we discuss our results and conclusion.

\begin{center}
\begin{table}[htb]
\begin{tabular}{|c|c|c|}
\hline
Parameters & NO & IO \\\hline
\hline
$ \frac{\Delta m_{21}^2}{10^{-5} \text{eV}^2}$ & $6.79-8.01$ & $6.79-8.01 $  \\
$ \frac{|\Delta m_{3l}^2|}{10^{-3} \text{eV}^2}$ & $2.427-2.625$ & $2.412-2.611 $  \\
$ \sin^2\theta_{12} $ &  $0.275-0.350 $ & $0.275-0.350 $  \\
$ \sin^2\theta_{23} $ & $0.418-0.627$ &  $0.423-0.629 $  \\
$\sin^2\theta_{13} $ & $0.02045-0.02439$ & $0.02068-0.02463 $  \\
$ \delta $ & $125^{\circ}-392^{\circ}$ & $196^{\circ}-360^{\circ}$  \\
\hline
\end{tabular}
\caption{Global fit $3\sigma$ values of neutrino oscillation parameters \cite{Esteban:2018azc}. Here $\Delta m_{3l}^2 \equiv \Delta m_{31}^2$ for NO and $\Delta m_{3l}^2 \equiv \Delta m_{32}^2$ for IO.}
\label{tab:data1}
\end{table}
\end{center}

\section{Texture Zero Mass Matrices}{\label{texture}}
A general $3\times3$ mass matrix can have nine complex or eighteen real independent parameters. For Majorana neutrinos, due to complex symmetric nature of the mass matrix, we have six independent complex parameters. For Dirac neutrinos, the mass matrix need not have any such symmetric properties in general and hence can have nine independent complex parameters. The previous works on Dirac neutrino textures \cite{Ahuja:2009jj, Liu:2012axa, Verma:2013cza, Fakay:2014nea, Sharma:2014wpa, Cebola:2015dwa, Ahuja:2016san, Singh:2018lao, Ahuja:2018fmw, Benavides:2020pjx} considered Hermitian mass matrices along with non-diagonal charged lepton mass matrices. In our work, we stick to the diagonal charged lepton basis for simplicity but consider both Hermitian as well as general structure of neutrino mass matrix. We discuss the two cases separately below.

\subsection{Case 1: Hermitian Mass Matrix}
Texture zero classification in this case is similar to that in Majorana neutrino scenario. Number of possibilities with $n$ zeros is $^6C_n$. While in case of Majorana neutrinos, only one possible two-zero texture and a few one-zero texture mass matrices are allowed from latest neutrino data \cite{Borgohain:2020now}, we expect similar or even more constrained scenario for Dirac neutrinos as well. Due to the absence of additional Majorana CP phases, the Dirac neutrino scenario for Hermitian mass matrix can be more constrained than the Majorana scenario as we discuss in details.  Therefore, we show the classification only for one-zero and two-zero textures of Dirac neutrinos in table \ref{10herm} and \ref{20herm} respectively. As expected, there are are $^6C_1=6$ and $^6C_2=15$ classes of possible one-zero and two-zero texture neutrino mass matrices. We do not show the classification of more than two-zero textures as they are expected to be disallowed in this case, from our knowledge about similar results in Majorana scenario.

\begin{table}[h!]\tiny
	\centering
	\begin{tabular}{| c|c| c| c| c| c| c|c| c||}
		\hline
		
		$A_1=\left(\begin{array}{ccc}
		0 & \times & \times\\
		\times& \times& \times\\
		\times & \times& \times
		\end{array}\right)$&	$A_2=\left(\begin{array}{ccc}
		\times & 0&\times \\
		0 & \times& \times\\
		\times& \times& \times
		\end{array}\right)$&$ A_3=\left(\begin{array}{ccc}
		\times & \times& 0\\
		\times& \times & \times\\
		0 & \times & \times
		\end{array}\right)$ & $A_4=\left(\begin{array}{ccc}
		\times & \times & \times\\
		\times& 0& \times\\
		\times & \times& \times
		\end{array}\right)$&$A_5=\left(\begin{array}{ccc}
		\times& \times &\times \\
		\times & \times& 0\\
		\times&0& \times
		\end{array}\right)$&$ A_6=\left(\begin{array}{ccc}
		\times & \times& \times\\
		\times& \times & \times\\
		\times & \times & 0
		\end{array}\right)$\\ \hline
	\end{tabular}
	\caption{1-0 texture neutrino mass matrices. The crosses "$\times$'' denote non-zero arbitrary elements.}
	\label{10herm}
\end{table}

\begin{table}[h!]\tiny
	\centering
	\begin{tabular}{| c|c| c| c| c| c| c|c| c||}
		\hline
		
		$B_1=\left(\begin{array}{ccc}
		0 & 0 & \times\\
		0& \times& \times\\
		\times & \times& \times
		\end{array}\right)$&	$B_2=\left(\begin{array}{ccc}
		0 & \times &0 \\
		\times & \times& \times\\
		0& \times& \times
		\end{array}\right)$&$B_3=\left(\begin{array}{ccc}
		\times & \times &0\\
		\times& 0& \times\\
		0 & \times& \times
		\end{array}\right)$ & $B_4=\left(\begin{array}{ccc}
		\times & 0&\times \\
		0 & \times& \times\\
		\times& \times& 0
		\end{array}\right)$&$B_5=\left(\begin{array}{ccc}
		\times & 0& \times \\
		0 & 0 & \times\\
		\times & \times & \times
		\end{array}\right)$&$ B_6=\left(\begin{array}{ccc}
		\times & \times& 0 \\
		\times & \times & \times\\
		0 & \times & 0
		\end{array}\right)$\\ \hline
		$B_7=\left(\begin{array}{ccc}
		\times & \times & \times\\
		\times& 0& \times\\
		\times & \times& 0
		\end{array}\right)$	&	$B_8=\left(\begin{array}{ccc}
		\times & \times &\times\\
		\times& 0& 0\\
		\times & 0& \times
		\end{array}\right)$&$B_9=\left(\begin{array}{ccc}
		\times & \times&\times \\
		\times & \times& 0\\
		\times& 0& 0
		\end{array}\right)$ &$B_{10}=\left(\begin{array}{ccc}
		0 & \times &\times\\
		\times& 0& \times\\
		\times & \times& \times
		\end{array}\right)$&$B_{11}=\left(\begin{array}{ccc}
		0 & \times&\times \\
		\times & \times& \times\\
		\times& \times& 0
		\end{array}\right)$&$B_{12}=\left(\begin{array}{ccc}
		0 & \times&\times \\
		\times & \times & 0\\
		\times & 0 & \times
		\end{array}\right)$\\ \hline
		$B_{13}=\left(\begin{array}{ccc}
		\times & 0 &0\\
		0& \times& \times\\
		0 & \times& \times
		\end{array}\right)$	&	$B_{14}=\left(\begin{array}{ccc}
		\times & 0&\times \\
		0 & \times& 0\\
		\times& 0& \times
		\end{array}\right)$&$ B_{15}=\left(\begin{array}{ccc}
		\times & \times& 0 \\
		\times & \times & 0\\
		0 & 0 & \times
		\end{array}\right)$ &&&\\ \hline
	\end{tabular}
	\caption{2-0 texture neutrino mass matrices. The crosses "$\times$'' denote non-zero arbitrary elements.}
	\label{20herm}
\end{table}
 
The Pontecorvo-Maki-Nakagawa-Sakata (PMNS) leptonic mixing matrix is given by 
\begin{equation}\label{eq7}
	U_{\text{PMNS}} = U^{\dagger}_{lL} U_{\nu L}
	\end{equation} 
where $U_{lL}, U_{\nu L}$ are diagonalising unitary matrices of charged lepton and light neutrino mass matrices, assuming them to be Hermitian. In the diagonal charged lepton basis we have $U_{lL} = \mathbb{1}$ and hence $U_{\rm PMNS}=U_{\nu L}$. The PMNS mixing matrix can be parametrised in terms of the leptonic mixing angles and phases as
	\begin{equation}\label{eq8}
	U_L=U_{\text{PMNS}}=\left[\begin{array}{ccc}
	c_{12}c_{13}&s_{12}c_{13}&s_{13}e^{-i\delta}\\
	-c_{23}s_{12}-s_{23}s_{13}c_{12}e^{i\delta}& c_{23}c_{12}-s_{23}s_{13}s_{12}e^{i\delta}&s_{23}c_{13}\\
	s_{23}s_{12}-c_{23}s_{13}c_{12}e^{i\delta}&-s_{23}c_{12}-c_{23}s_{13}s_{12}e^{i\delta}&c_{23}c_{13}
	\end{array}\right]P
	\end{equation}
	where $c_{ij} = \cos{\theta_{ij}}, \; s_{ij} = \sin{\theta_{ij}}$ and $\delta$ is the leptonic Dirac CP phase. The diagonal matrix $P$ contains the Majorana CP phases if neutrinos are of Majorana nature. For Dirac neutrinos, these Majorana phases are vanishing and hence $P = \mathbb{1}$. Thus, the light neutrino mass matrix can be constructed as 
	\begin{equation}\label{eq6}
	M_\nu= U_{\nu L} {M_\nu}^{(\rm diag)} U^{\dagger}_{\nu L}= U_{\rm PMNS}{M_\nu}^{(\rm diag)} {U^{\dagger}_{\rm PMNS}}
	\end{equation}
The diagonal light neutrino mass matrix is denoted by ${M_\nu}^{(\rm diag)}= \textrm{diag}(m_1,m_2,m_3)$ where the light neutrino masses can follow either normal ordering (NO) or inverted ordering (IO). For NO, the three neutrino mass eigenvalues can be written as 
$$M^{\text{diag}}_{\nu}
= \text{diag}(m_1, \sqrt{m^2_1+\Delta m_{21}^2}, \sqrt{m_1^2+\Delta m_{31}^2})$$ while for IO, they can be written as 
$$M^{\text{diag}}_{\nu} = \text{diag}(\sqrt{m_3^2+\Delta m_{23}^2-\Delta m_{21}^2}, \sqrt{m_3^2+\Delta m_{23}^2}, m_3)$$ 
The analytical expressions of the elements of this mass matrix are given in Appendix \ref{appen1}. 

It can be seen from the expressions for mass matrix elements given in Appendix \ref{appen1} that the diagonal entries are real whereas the off-diagonal elements are complex, as expected. While we check the validity of texture zeros numerically, it is simple enough to make some estimates in Hermitian case. Clearly, any texture zero mass matrix having $m_{ee}=c_{13}^2 \left(c_{12}^2 m_1+m_2 s_{12}^2\right)+m_3 s_{13}^2 =0$ will be disallowed as there is no way to make this element vanishing while being in agreement with light neutrino data. Similarly, $m_{e\mu}=0$ will lead to 
$$ c_{12} c_{23} \left(m_2-m_1\right) s_{12}-\cos{\delta} s_{13} s_{23}
   \left(c_{12}^2 m_1+m_2 s_{12}^2-m_3\right) = 0 $$
   $$\sin{\delta} s_{13} s_{23}
   \left(c_{12}^2 m_1+m_2 s_{12}^2-m_3\right) =0 $$
   While the second condition implies $\delta = 0, \pi, 2\pi$, the first condition can not satisfy light neutrino data. For example, taking $\delta = 0, 2\pi$, it leads to  
   $$c_{12} c_{23} \left(m_2-m_1\right) s_{12}=s_{13} s_{23}
   \left(c_{12}^2 m_1+m_2 s_{12}^2-m_3\right) $$
   which can be true only for IO of light neutrino mass. Assuming $m_3 \ll m_1 <m_2$, we get, by using best fit values of mixing angles, $m_2 \approx 1.42 m_1$ which does not satisfy the solar mass splitting data for this hierarchical limit. Similarly, $\delta = \pi$ which can be true for NO only, leads to $m_3/m_2 \approx 2.67$ which falls outside the $3\sigma$ range $m_3/m_2 \approx \sqrt{\Delta m^2_{31}/\Delta m^2_{21}} \in 5.50-6.22$ assuming a hierarchical limit $m_1 \ll m_2 <m_3$. Similarly, one can explain the validity of other one-zero textures of Hermitian case. By performing a numerical calculation, varying all known parameters in $3\sigma$ range, we show that both one-zero and two-zero textures are disallowed if Dirac neutrino mass matrix is Hermitian and charged lepton mass matrix is diagonal.

\section{Case 2: General Mass Matrix}
\label{sec3a}

\begin{table}[h!]\tiny
	\centering
\begin{tabular}{| c|c| c| c| c| c| c|c| c||}
		\hline
		
		$A_1=\left(\begin{array}{ccc}
		0 & \times &\times\\
		\times& \times& \times\\
		\times & \times& \times
		\end{array}\right)$&	$A_2=\left(\begin{array}{ccc}
		\times & 0&\times \\
		\times & \times& \times\\
		\times& \times& \times
		\end{array}\right)$&$ A_3=\left(\begin{array}{ccc}
		\times & \times& 0\\
		\times& \times & \times\\
		\times & \times & \times
		\end{array}\right)$ & $A_4=\left(\begin{array}{ccc}
		\times & \times & \times\\
		0& \times& \times\\
		\times & \times& \times
		\end{array}\right)$&$A_5=\left(\begin{array}{ccc}
		\times& \times &\times \\
		\times & 0& \times\\
		\times&\times& \times
		\end{array}\right)$&$ A_6=\left(\begin{array}{ccc}
		\times & \times& \times\\
		\times& \times & 0\\
		\times & \times & \times
		\end{array}\right)$\\ \hline
		$A_7=\left(\begin{array}{ccc}
		\times & \times & \times\\
		\times& \times& \times\\
		0 & \times& \times
		\end{array}\right)$	&	$A_8=\left(\begin{array}{ccc}
		\times& \times &\times \\
		\times & \times& \times\\
		\times& 0& \times
		\end{array}\right)$&$A_9=\left(\begin{array}{ccc}
		\times & \times& \times\\
		\times& \times & \times\\
		\times & \times & 0
		\end{array}\right)$ &&&\\ \hline
		\end{tabular}
	\caption{1-0 texture neutrino mass matrices. The crosses "$\times$'' denote non-zero arbitrary elements}
	\label{10nonherm}
\end{table}

\begin{table}[h!]\tiny
	\centering
	\begin{tabular}{| c|c| c| c| c| c| c|c| c||}
		\hline
		
		$B_1=\left(\begin{array}{ccc}
		0 & 0 & \times\\
		\times& \times& \times\\
		\times & \times& \times
		\end{array}\right)$&	$B_2=\left(\begin{array}{ccc}
		0 & \times &0 \\
		\times & \times& \times\\
		\times& \times& \times
		\end{array}\right)$&$B_3=\left(\begin{array}{ccc}
		0 & \times &\times \\
		0 & \times& \times\\
		\times& \times& \times
		\end{array}\right)$ & $B_4=\left(\begin{array}{ccc}
		0 & \times &\times \\
		\times & 0& \times\\
		\times& \times& \times
		\end{array}\right)$&$B_5=\left(\begin{array}{ccc}
		0 & \times & \times\\
		\times& \times& 0\\
		\times & \times& \times
		\end{array}\right)$&$B_6=\left(\begin{array}{ccc}
		0 & \times & \times \\
		\times & \times& \times\\
		0& \times& \times
		\end{array}\right)$\\ \hline
		$B_7=\left(\begin{array}{ccc}
		0 & \times &\times \\
		\times& \times& \times\\
		\times& 0& \times
		\end{array}\right)$	&	$B_8=\left(\begin{array}{ccc}
		0 & \times &\times \\
		\times &  \times& \times\\
		\times& \times& 0
		\end{array}\right)$&$B_9=\left(\begin{array}{ccc}
		\times & 0 & 0\\
		\times& \times& \times\\
		\times & \times& \times
		\end{array}\right)$ &$B_{10}=\left(\begin{array}{ccc}
		\times & 0 & \times \\
		0 & \times& \times\\
		\times& \times& \times
		\end{array}\right)$&$B_{11}=\left(\begin{array}{ccc}
		\times & 0 &\times \\
		\times& 0& \times\\
		\times& \times& \times
		\end{array}\right)$&$B_{12}=\left(\begin{array}{ccc}
		\times & 0 &\times \\
		\times &  \times& 0\\
		\times& \times& \times
		\end{array}\right)$\\
\hline
		\end{tabular}
	\caption{2-0 texture neutrino mass matrices. The crosses "$\times$'' denote non-zero arbitrary elements}
	\label{20nonherm}
\end{table}

\begin{table}[h!]\tiny
	\centering
	\begin{tabular}{| c|c| c| c| c| c| c|c| c||}
		\hline
		
		$C_{1}=\left(\begin{array}{ccc}
		0 & 0 & 0\\
		\times& \times& \times\\
		\times & \times& \times
		\end{array}\right)$&	$C_{2}=\left(\begin{array}{ccc}
		0 & 0& \times\\
		0& \times& \times\\
		\times& \times& \times
		\end{array}\right)$&$C_{3}=\left(\begin{array}{ccc}
		0 & 0 &\times\\
		\times& 0& \times\\
		\times& \times& \times
		\end{array}\right)$ & $C_{4}=\left(\begin{array}{ccc}
		0 &0 &\times \\
		\times&  \times& 0\\
		\times& \times& \times
		\end{array}\right)$&$C_{5}=\left(\begin{array}{ccc}
		0 &0 & \times\\
		\times& \times& \times\\
		0 & \times& \times
		\end{array}\right)$&$C_{6}=\left(\begin{array}{ccc}
		0 & 0& \times\\
		\times& \times& \times\\
		\times& 0& \times
		\end{array}\right)$\\ \hline
		$C_{7}=\left(\begin{array}{ccc}
		0 & 0 &\times\\
		\times& \times& \times\\
		\times& \times& 0
		\end{array}\right)$	&	$C_{8}=\left(\begin{array}{ccc}
		0 & \times &0 \\
		0&  \times& \times\\
		\times& \times& \times
		\end{array}\right)$&$C_{9}=\left(\begin{array}{ccc}
		0 &\times & 0\\
		\times& \times& 0\\
		\times & \times& \times
		\end{array}\right)$ &$C_{10}=\left(\begin{array}{ccc}
		0 &\times & 0\\
		\times& 0& \times\\
		\times & \times& \times
		\end{array}\right)$&$C_{11}=\left(\begin{array}{ccc}
		0 &\times & 0\\
		\times& \times& \times\\
		0 & \times& \times
		\end{array}\right)$&$C_{12}=\left(\begin{array}{ccc}
		0 & \times &0 \\
		\times&  \times& \times\\
		\times& 0& \times
		\end{array}\right)$\\
\hline									
	\end{tabular}
	\caption{3-0 texture neutrino mass matrices. The crosses "$\times$'' denote non-zero arbitrary elements.}
	\label{30nonherm}
\end{table}

\begin{table}[h!]\tiny
	\centering
	\begin{tabular}{| c|c| c| c| c| c| c|c| c||}
		\hline
		
		$D_{1}=\left(\begin{array}{ccc}
		0 & 0 & 0\\
		\times& \times& 0\\
		\times & \times& \times
		\end{array}\right)$&	$D_{2}=\left(\begin{array}{ccc}
		\times & 0& 0\\
		0& \times& 0\\
		\times& \times& \times
		\end{array}\right)$&$D_{3}=\left(\begin{array}{ccc}
		\times & 0 &0\\
		\times& 0& 0\\
		\times& \times& \times
		\end{array}\right)$ & $D_{4}=\left(\begin{array}{ccc}
		\times &0 &0 \\
		\times&  \times& 0\\
		0& \times& \times
		\end{array}\right)$&$D_{5}=\left(\begin{array}{ccc}
		\times &0 & 0\\
		\times& \times& 0\\
		\times & 0& \times
		\end{array}\right)$&$D_{6}=\left(\begin{array}{ccc}
		\times & 0& 0\\
		\times& \times& 0\\
		\times& \times& 0
		\end{array}\right)$\\ \hline
		$D_{7}=\left(\begin{array}{ccc}
		0 & 0 &0\\
		\times& \times& \times\\
		\times& 0&\times
		\end{array}\right)$	&	$D_{8}=\left(\begin{array}{ccc}
		\times  & 0 &0 \\
		0&  \times& \times\\
		\times& 0& \times
		\end{array}\right)$&$D_{9}=\left(\begin{array}{ccc}
		\times  &0 & 0\\
		\times& 0&\times\\
		\times & 0& \times
		\end{array}\right)$ &$D_{10}=\left(\begin{array}{ccc}
		\times &0 & 0\\
		\times& \times& \times\\
		0 & 0& \times
		\end{array}\right)$&$D_{11}=\left(\begin{array}{ccc}
		\times &0 & 0\\
		\times& \times& \times\\
		\times & 0& 0
		\end{array}\right)$&$D_{12}=\left(\begin{array}{ccc}
		0 & 0 &0 \\
		\times&  \times& \times\\
		\times& \times&0
		\end{array}\right)$\\
\hline																
	\end{tabular}
	\caption{4-0 texture neutrino mass matrices.}
	\label{40nonherm}
\end{table}

\begin{table}[h!]\tiny
	\centering
	\begin{tabular}{| c|c| c| c| c| c| c|c| c||}
		\hline
		
		$E_{1}=\left(\begin{array}{ccc}
		\times & \times & \times\\
		0& 0& \times\\
		0 & 0& 0
		\end{array}\right)$&	$E_{2}=\left(\begin{array}{ccc}
		0 & \times& \times\\
		\times& 0& \times\\
		0& 0& 0
		\end{array}\right)$&$E_{3}=\left(\begin{array}{ccc}
		0 & \times &\times\\
		0& \times& \times\\
	0& 0& 0
		\end{array}\right)$ & $E_{4}=\left(\begin{array}{ccc}
		0 &\times &\times \\
		0&  0& \times\\
		\times& 0& 0
		\end{array}\right)$&$E_{5}=\left(\begin{array}{ccc}
		0 &\times & \times\\
		0 &0& \times\\
		0 & \times& 0
		\end{array}\right)$&$E_{6}=\left(\begin{array}{ccc}
		0 &\times & \times\\
		0& 0& \times\\
		0& 0& \times
		\end{array}\right)$\\ \hline
		$E_{7}=\left(\begin{array}{ccc}
		\times & \times& \times\\
0& 0& 0\\
	0& \times&0
		\end{array}\right)$	&	$E_{8}=\left(\begin{array}{ccc}
			0 & \times& \times\\
		\times& 0& 0\\
		0& \times&0 
		\end{array}\right)$&$E_{9}=\left(\begin{array}{ccc}
			0 & \times& \times\\
		0& \times& 0\\
		0& \times& 0
		\end{array}\right)$ &$E_{10}=\left(\begin{array}{ccc}
			0 & \times& \times\\
		0& 0& 0\\
		\times& \times& 0
		\end{array}\right)$&$E_{11}=\left(\begin{array}{ccc}
		0 & \times& \times\\
		0& 0& 0\\
		0& \times& \times
		\end{array}\right)$&$E_{12}=\left(\begin{array}{ccc}
		\times & \times& \times\\
	0& 0& 0\\
	0& 0& \times
		\end{array}\right)$\\
\hline																
	\end{tabular}
	\caption{5-0 texture neutrino mass matrices.}
	\label{50nonherm}
\end{table}

\begin{table}[h]

	\begin{tabular}{|c|c |  c| }
		\hline
		Class&NON-HERMITIAN	   \\ \hline\hline
		$A_1$(NO/IO)&$ \checkmark(\checkmark)$ \\ \hline
		$A_2$(NO/IO)&$ \checkmark(\checkmark)$ \\ \hline
		$A_3$(NO/IO)&$ \checkmark (\checkmark)$  \\ \hline
		$A_4$(NO/IO)&$ \checkmark (\checkmark)$  \\ \hline
		$A_5$(NO/IO)&$ \checkmark (\checkmark)$     \\ \hline
		$A_6$(NO/IO)&$ \checkmark (\checkmark)$ \\ \hline
		$A_7$(NO/IO)&$ \checkmark (\checkmark)$ \\ \hline
		$A_8$(NO/IO)&$ \checkmark (\checkmark) $      \\ \hline
		$A_9$(NO/IO)& $ \checkmark (\checkmark)$   \\ \hline	
		\end{tabular}
	
	\caption{Summary of allowed one-zero textures based on current experimental bounds.}\label{t1}
\end{table}

	\begin{table}[!htb]
	
	\begin{minipage}{.5\linewidth}
		
		\centering
		\begin{tabular}{|c|c | }
			\hline
			Class&	NON-HERMITIAN \\ \hline\hline
			$B_1$(NO/IO)&$ \checkmark (\checkmark)$	 \\ \hline
			$B_2$(NO/IO)&$ \checkmark (\checkmark)$	 \\ \hline
			$B_3$(NO/IO)&$ \checkmark (\checkmark)$ \\ \hline
			$B_4$(NO/IO)&$ \checkmark (\checkmark)$ 	 \\ \hline
			$B_5$(NO/IO)&$ \checkmark (\checkmark)$	   \\ \hline
			$B_6$(NO/IO)&$ \checkmark (\checkmark)$   \\ \hline
			$B_7$(NO/IO)&$ \checkmark (\checkmark)$  \\ \hline
			$B_8$(NO/IO)&$ \checkmark (\checkmark)$   \\ \hline
			$B_9$(NO/IO)& $ \checkmark (\times) $\\ \hline	
			$B_{10}$(NO/IO)&$ \checkmark (\checkmark)$ \\ \hline	
			$B_{11}$(NO/IO)&$ \checkmark (\checkmark)$   \\ \hline
			$B_{12}$(NO/IO)&$ \checkmark (\checkmark)$   \\ \hline	
			$B_{13}$(NO/IO)&$ \checkmark (\checkmark)$   \\ \hline
			$B_{14}$(NO/IO)&$ \checkmark (\checkmark)$   \\ \hline
			$B_{15}$(NO/IO)&$ \checkmark (\checkmark)$   \\ \hline
			$B_{16}$(NO/IO)&$ \checkmark (\checkmark)$  \\ \hline
			$B_{17}$(NO/IO)&$ \checkmark (\checkmark)$  \\ \hline
			$B_{18}$(NO/IO)&$ \checkmark (\checkmark)$  \\ \hline

		\end{tabular}
	\end{minipage}%
	\begin{minipage}{.5\linewidth}
		\centering
		
		\begin{tabular}{|c|c | }
			\hline
			Class&NON-HERMITIAN  \\ \hline\hline
			$B_{19}$(NO/IO)&$ \checkmark (\checkmark)$  \\ \hline
			$B_{20}$(NO/IO)&$ \checkmark (\checkmark)$	  \\ \hline
			$B_{21}$(NO/IO)&$ \checkmark(\checkmark)$ \\ \hline
			$B_{22}$(NO/IO)&$ \checkmark (\checkmark)$\\ \hline
			$B_{23}$(NO/IO)&$ \checkmark (\checkmark)$  \\ \hline
			$B_{24}$(NO/IO)&$  \checkmark (\checkmark)$ 	 \\ \hline
			$B_{25}$(NO/IO)&$ \checkmark (\checkmark)$  \\ \hline
			$B_{26}$(NO/IO)&$ \checkmark (\checkmark)$  \\ \hline
			$B_{27}$(NO/IO)&$ \checkmark (\checkmark)$  \\ \hline
			$B_{28}$(NO/IO)&$ \checkmark (\checkmark)$ \\ \hline	
			$B_{29}$(NO/IO)&$ \checkmark (\checkmark)$  \\ \hline
			$B_{30}$(NO/IO)&$ \checkmark (\checkmark)$  \\ \hline
			$B_{31}$(NO/IO)&$ \checkmark(\checkmark)$   \\ \hline
			$B_{32}$(NO/IO)&$ \checkmark (\checkmark)$ \\ \hline
			$B_{33}$(NO/IO)&$ \checkmark (\checkmark)$  \\ \hline
			$B_{34}$(NO/IO)&$ \checkmark (\checkmark)$   \\ \hline
			$B_{35}$(NO/IO)&$ \checkmark (\checkmark)$   \\ \hline
			$B_{36}$(NO/IO)&$ \checkmark (\checkmark)$  \\ \hline

		\end{tabular}
	\end{minipage} 
	\caption{Summary of allowed and disallowed two-zero textures based on current experimental bounds.}\label{t2}
\end{table}

	\begin{table}[!htb]
	
	\begin{minipage}{.5\linewidth}
		
		\centering
		\begin{tabular}{|c|c | }
			\hline
			Class&	NON-HERMITIAN \\ \hline\hline
			$C_1$(NO/IO)&$ \times (\times)$	 \\ \hline
			$C_2$(NO/IO)&$ \checkmark (\checkmark)$	 \\ \hline
			$C_3$(NO/IO)&$ \checkmark (\checkmark)$ \\ \hline
			$C_4$(NO/IO)&$ \times (\times)$ 	 \\ \hline
			$C_5$(NO/IO)&$ \checkmark (\checkmark)$	   \\ \hline
			$C_6$(NO/IO)&$ \checkmark (\checkmark)$   \\ \hline
			$C_7$(NO/IO)&$ \times (\times)$  \\ \hline
			$C_8$(NO/IO)&$ \checkmark (\checkmark)$   \\ \hline
			$C_9$(NO/IO)& $ \checkmark (\checkmark) $\\ \hline	
			$C_{10}$(NO/IO)&$ \times(\times)$ \\ \hline	
			$C_{11}$(NO/IO)&$ \checkmark (\checkmark)$   \\ \hline
			$C_{12}$(NO/IO)&$ \times (\times)$   \\ \hline	
			$C_{13}$(NO/IO)&$ \checkmark (\checkmark)$   \\ \hline
			$C_{14}$(NO/IO)&$ \checkmark (\checkmark)$   \\ \hline
			$C_{15}$(NO/IO)&$ \checkmark (\checkmark)$   \\ \hline
			$C_{16}$(NO/IO)&$ \checkmark (\times)$  \\ \hline
			$C_{17}$(NO/IO)&$ \checkmark (\checkmark)$  \\ \hline
			$C_{18}$(NO/IO)&$ \checkmark (\checkmark)$  \\ \hline
			$C_{19}$(NO/IO)&$ \times (\times)$  \\ \hline
			$C_{20}$(NO/IO)&$ \checkmark (\checkmark)$	  \\ \hline
			$C_{21}$(NO/IO)&$ \checkmark(\times)$ \\ \hline

		\end{tabular}
	\end{minipage}%
	\begin{minipage}{.5\linewidth}
		\centering
		
		\begin{tabular}{|c|c | }
			\hline
			Class&NON-HERMITIAN  \\ \hline\hline
			$C_{22}$(NO/IO)&$ \checkmark (\checkmark)$\\ \hline
			$C_{23}$(NO/IO)&$ \checkmark (\checkmark)$  \\ \hline
			$C_{24}$(NO/IO)&$  \checkmark (\checkmark)$ 	 \\ \hline
			$C_{25}$(NO/IO)&$ \checkmark (\times)$  \\ \hline
			$C_{26}$(NO/IO)&$ \checkmark (\checkmark)$  \\ \hline
			$C_{27}$(NO/IO)&$ \checkmark (\checkmark)$  \\ \hline
			$C_{28}$(NO/IO)&$ \times (\times)$ \\ \hline
			$C_{29}$(NO/IO)&$ \times (\times)$  \\ \hline
			$C_{30}$(NO/IO)&$ \checkmark (\times)$  \\ \hline
			$C_{31}$(NO/IO)&$ \checkmark(\times)$   \\ \hline
			$C_{32}$(NO/IO)&$ \times (\times)$ \\ \hline
			$C_{33}$(NO/IO)&$ \checkmark (\times)$  \\ \hline
			$C_{34}$(NO/IO)&$ \checkmark (\times)$   \\ \hline
			$C_{35}$(NO/IO)&$ \checkmark (\checkmark)$   \\ \hline
			$C_{36}$(NO/IO)&$ \times (\times)$  \\ \hline
			$C_{37}$(NO/IO)&$ \checkmark (\checkmark)$  \\ \hline
			$C_{38}$(NO/IO)&$ \checkmark (\checkmark)$  \\ \hline
			$C_{39}$(NO/IO)&$ \checkmark(\checkmark)$   \\ \hline
			$C_{40}$(NO/IO)&$ \checkmark (\checkmark)$ \\ \hline
			$C_{41}$(NO/IO)&$ \checkmark (\checkmark)$  \\ \hline
			$C_{42}$(NO/IO)&$ \checkmark (\times)$   \\ \hline

		\end{tabular}
	\end{minipage} 
	\caption{Summary of allowed and disallowed three-zero textures based on current experimental bounds.}\label{t22}
\end{table}

	\begin{table}[!htb]
	
	\begin{minipage}{.5\linewidth}
		
		\centering
		\begin{tabular}{|c|c | }
			\hline
			Class&	NON-HERMITIAN \\ \hline\hline
			$C_{43}$(NO/IO)&$ \checkmark (\checkmark)$   \\ \hline
			$C_{44}$(NO/IO)&$ \checkmark (\checkmark)$  \\ \hline
			$C_{45}$(NO/IO)&$ \checkmark (\checkmark)$	 \\ \hline
			$C_{46}$(NO/IO)&$ \checkmark (\times)$ \\ \hline
			$C_{47}$(NO/IO)&$ \checkmark (\checkmark)$ 	 \\ \hline
			$C_{48}$(NO/IO)&$ \times (\times)$	   \\ \hline
			$C_{49}$(NO/IO)&$ \checkmark (\checkmark)$   \\ \hline
			$C_{50}$(NO/IO)&$ \times (\times)$  \\ \hline
			$C_{51}$(NO/IO)&$ \checkmark (\checkmark)$   \\ \hline
			$C_{52}$(NO/IO)& $ \checkmark (\times) $\\ \hline	
			$C_{53}$(NO/IO)&$ \checkmark (\checkmark)$ \\ \hline	
			$C_{54}$(NO/IO)&$ \checkmark (\checkmark)$   \\ \hline
			$C_{55}$(NO/IO)&$ \times (\times)$   \\ \hline	
			$C_{56}$(NO/IO)&$ \checkmark (\checkmark)$   \\ \hline
			$C_{57}$(NO/IO)&$ \checkmark (\checkmark)$   \\ \hline
			$C_{58}$(NO/IO)&$ \times (\times)$   \\ \hline
			$C_{59}$(NO/IO)&$ \checkmark (\checkmark)$  \\ \hline
			$C_{60}$(NO/IO)&$ \checkmark (\checkmark)$  \\ \hline
			$C_{61}$(NO/IO)&$ \checkmark (\checkmark)$  \\ \hline
			$C_{62}$(NO/IO)&$ \checkmark (\checkmark)$  \\ \hline
			$C_{63}$(NO/IO)&$ \times (\times)$	  \\ \hline
		
		\end{tabular}
	\end{minipage}%
	\begin{minipage}{.5\linewidth}
		\centering
		
		\begin{tabular}{|c|c | }
			\hline
			Class&NON-HERMITIAN  \\ \hline\hline
			$C_{64}$(NO/IO)&$ \checkmark(\checkmark)$ \\ \hline
			$C_{65}$(NO/IO)&$ \checkmark (\checkmark)$\\ \hline
			$C_{66}$(NO/IO)&$ \times (\times)$  \\ \hline
			$C_{67}$(NO/IO)&$  \checkmark (\checkmark)$ 	 \\ \hline
			$C_{68}$(NO/IO)&$ \checkmark (\checkmark)$  \\ \hline
			$C_{69}$(NO/IO)&$ \times (\times)$  \\ \hline
			$C_{70}$(NO/IO)&$ \checkmark (\checkmark)$  \\ \hline
			$C_{71}$(NO/IO)&$ \times (\times)$ \\ \hline	
			$C_{72}$(NO/IO)&$ \checkmark (\checkmark)$ \\ \hline
			$C_{73}$(NO/IO)&$ \checkmark (\checkmark)$  \\ \hline
			$C_{74}$(NO/IO)&$ \checkmark (\checkmark)$  \\ \hline
			$C_{75}$(NO/IO)&$ \times (\times)$  \\ \hline
			$C_{76}$(NO/IO)&$ \times (\times)$	  \\ \hline
			$C_{77}$(NO/IO)&$ \checkmark(\checkmark)$ \\ \hline
			$C_{78}$(NO/IO)&$ \checkmark (\checkmark)$\\ \hline
			$C_{79}$(NO/IO)&$ \checkmark (\checkmark)$  \\ \hline
			$C_{80}$(NO/IO)&$  \times(\times)$ 	 \\ \hline
			$C_{81}$(NO/IO)&$ \checkmark (\checkmark)$  \\ \hline
			$C_{82}$(NO/IO)&$ \times (\times)$  \\ \hline
			$C_{83}$(NO/IO)&$ \checkmark (\checkmark)$  \\ \hline
			$C_{84}$(NO/IO)&$ \checkmark (\checkmark)$ \\ \hline

		\end{tabular}
	\end{minipage} 
	\caption{Summary of allowed and disallowed three-zero textures based on current experimental bounds.}\label{t222}
\end{table}

	\begin{table}[!htb]
\centering
	\begin{tabular}{|c|c | }
			\hline
			 Allowed Classes (NO)& Allowed Classes (IO)  \\ \hline\hline
			
			$D_{16},D_{34},D_{76},D_{83}$&$D_{16},D_{76}$\\ 
			$D_{96},D_{98}$&\\\hline	
	 
	\end{tabular}
 
	\caption{Summary of allowed four-zero textures based on current experimental bounds.}\label{t3}
\end{table}

For general Dirac neutrino mass matrices, the $3\times 3$ matrix can have nine independent complex elements. Therefore, the number of possible n-zero mass matrices will be $^9C_n$. Clearly, such a general case not only introduces more free parameters but also allows many more possible texture zeros in Dirac neutrino mass matrix. For example, there are nine one-zero and thirty six two-zero textures compared to six one-zero and fifteen two-zero textures in Hermitian scenario. Similarly, ruling out three-zero textures in Hermitian scenario (or in Majorana neutrino scenario) does not necessarily rule out such higher zero texture possibilities for general mass matrix. We, therefore list all such possibilities up to four-zero textures in table \ref{10nonherm}, \ref{20nonherm}, \ref{30nonherm}, \ref{40nonherm} respectively. Since we found (as discussed below) all four-zero textures to be disallowed this automatically rules out higher texture zeros and hence we do not show the classification beyond four-zero textures.

Now, a general complex Dirac neutrino mass matrix can not be diagonalised by a simple unitary transformation. For this, we need a bi-unitary transformation as follows
	\begin{equation}\label{eq6}
	M_\nu= U_{\nu L} {M_\nu}^{(\rm diag)} U^{\dagger}_{\nu R}
	\end{equation}
where $U_{\nu L}, U_{\nu R}$ are two unitary matrices with $U_{\nu L}$ being the same unitary matrix that appears in PMNS leptonic mixing matrix mentioned before. In the diagonal charged lepton basis, one can again write $U_{\nu L} = U_{\rm PMNS}$. The other unitary matrix $U^{\dagger}_{\nu R}$ is arbitrary and can be parametrised in terms of arbitrary rotation angles and phases. In general, a $3\times3$ unitary matrix has three angles and six phases. From the leptonic mixing matrix, five phases can be rotated away by using the freedom of redefining the fields. However, since such freedom of rotating away phases have already been utilised, it is no longer possible, in general, to rotate away any of the phases in $U_{\nu R}$. This is particularly true in left-right symmetric models (LRSM) with Dirac neutrinos \cite{Pati:1974yy, Mohapatra:1974hk, Mohapatra:1974gc, Senjanovic:1975rk, Senjanovic:1978ev, Herczeg:1985cx, Babu:1988yq, Ma:1989tz, Ma:2016mwh, Borah:2016lrl, Borah:2016hqn, Borah:2017leo, Ma:2017kgb, Chavez:2019yal, Borah:2020boy}. We can parametrise it as \cite{Senjanovic:2015yea}
\begin{equation}
U_{\nu R} = {\rm diag} \left (e^{i \omega_1}, e^{i \omega_2}, e^{i \omega_3} \right) V_R (\theta^R_{ij}, \delta_R) {\rm diag} \left (e^{i \omega_4}, e^{i \omega_5}, 1 \right) 
\end{equation}
where $V_R$ can be written in a way similar to the PMNS matrix as 
	\begin{equation}\label{eq8}
	V_R (\theta^R_{ij}, \delta_R)=\left[\begin{array}{ccc}
	cr_{12}cr_{13}&sr_{12}cr_{13}&sr_{13}e^{-i\delta_R}\\
	-cr_{23}sr_{12}-sr_{23}sr_{13}cr_{12}e^{i\delta_R}& cr_{23}cr_{12}-sr_{23}sr_{13}sr_{12}e^{i\delta_R}&sr_{23}cr_{13}\\
	sr_{23}sr_{12}-cr_{23}sr_{13}cr_{12}e^{i\delta_R}&-sr_{23}cr_{12}-cr_{23}sr_{13}sr_{12}e^{i\delta_R}&cr_{23}cr_{13}
	\end{array}\right]
	\end{equation}
	where $cr_{ij} = \cos{\theta^R_{ij}}, \; sr_{ij} = \sin{\theta^R_{ij}}$ and $\delta_R$ is a CP phase similar to $\omega_1, \omega_2, \omega_3, \omega_4, \omega_5$. Using this parametrisation, one can derive the analytical form of Dirac neutrino mass matrix. The analytical form of mass matrix elements are given in Appendix \ref{appen2}.
	
As can be seen from the expressions of mass matrix elements Appendix \ref{appen2}, there are many more free parameters entering into them compared to Hermitian case discussed earlier. However, the newly introduced nine parameters (three new mixing angles and six new phases) do not occur separately in the expressions, but appear in certain combinations. This restricts the number of allowed texture zeros as we discuss below. Since the new phases appear in certain combinations only, we still do not have enough freedom to have more than four zeros in light neutrino mass matrix while being consistent with light neutrino data. Apart from such predictive nature in deciding whether a texture is allowed or not, we also get interesting predictions for light neutrino parameters in allowed textures as we discuss below.

Using the expressions given in Appendix \ref{appen2}, texture zero conditions say, $m_{ee}=0, m_{e \mu}=0$ lead to
\begin{equation}
 e^{i\omega_5} U_{11} V^*_{11}m_1+e^{i \omega_4} (U_{12}V^*_{12}m_2+e^{i \omega_5} U_{13} V^*_{13} m_3)=0,
\end{equation}
\begin{equation}
 e^{i\omega_5} U_{11} V^*_{21}m_1+e^{i \omega_4} (U_{12}V^*_{22}m_2+e^{i \omega_5} U_{13} V^*_{23} m_3) =0.
\end{equation}
Upon some simplifications, we can derive
\begin{equation}
\frac{m_2}{m_1} e^{i(\omega_4-\omega_5)} = \frac{U_{11}(V^*_{21}V^*_{13}-V^*_{11} V^*_{23})}{U_{12}(V^*_{12} V^*_{23}-V^*_{22} V^*_{13})} \implies \bigg \lvert \frac{m_2}{m_1} \bigg \rvert = \frac{ \lvert U_{11}(V^*_{21}V^*_{13}-V^*_{11} V^*_{23}) \rvert}{\lvert U_{12}(V^*_{12} V^*_{23}-V^*_{22} V^*_{13}) \rvert}, 
\end{equation}
\begin{equation}
\frac{m_3}{m_1} e^{i(\omega_4)} = \frac{U_{11}(V^*_{21}V^*_{12}-V^*_{11} V^*_{22})}{U_{13}(V^*_{13} V^*_{22}-V^*_{23} V^*_{12})} \implies \bigg \lvert \frac{m_3}{m_1} \bigg \rvert = \frac{\lvert U_{11}(V^*_{21}V^*_{12}-V^*_{11} V^*_{22})\rvert}{\lvert U_{13}(V^*_{13} V^*_{22}-V^*_{23} V^*_{12})\rvert }. 
\end{equation}
We can further simplify it if we assume some specific structure of $V_R$. For example, if we take $V_{13} \approx 0$, we can find 
\begin{equation}
\frac{m_2}{m_1} = \frac{U_{11}}{U_{12}} \cot{(\theta^R_{12})}, \; \frac{m_3}{m_1} = \frac{U_{11}}{\lvert U_{13} \rvert} \cot{(\theta^R_{23})} {\rm cosec} {(\theta^R_{12})}.
\label{twozero1}
\end{equation}
These relations can lead to interesting correlations between different mixing angles and lightest neutrino mass as we discuss in upcoming section. While these relations do not involve the leptonic Dirac CP phase, it appears in other types of texture zero conditions. For example, $m_{\mu e}=0, m_{\mu \mu}=0$ leads to
\begin{equation}
\frac{m_2}{m_1} = \bigg \lvert \frac{U_{21}}{U_{22}} \bigg \rvert \cot{(\theta^R_{12})}, \; \frac{m_3}{m_1} = \bigg \lvert \frac{U_{21}}{U_{23}} \bigg \rvert \cot{(\theta^R_{23})} {\rm cosec} {(\theta^R_{12})}
\label{twozero2}
\end{equation}
where dependence on leptonic Dirac CP phase $\delta$ will be explicit via $U_{21}, U_{22}$.

Similarly, for more texture zeros, one can find relations between neutrino parameters. Clearly, three of the additional CP phases namely, $\omega_1, \omega_2, \omega_3$ do not appear in the texture zero conditions. Therefore, only three mixing angles in $V_R$, three remaining CP phases appear along with light neutrino parameters in the texture zero conditions. Since we have two unknowns in light neutrino sector namely, Dirac CP phase and the lightest neutrino mass, effectively we have a total of eight free parameters appearing in the texture zero conditions. Thus, any texture more than four zeros lead to too much restrictions on known parameters as well and as expected, we find all five-zero textures disallowed in our numerical analysis discussed below.
\begin{figure}[h]
	\includegraphics[width=0.47\textwidth]{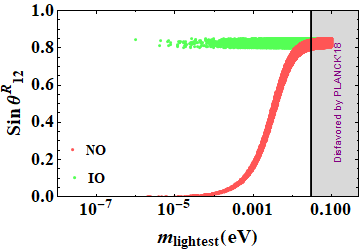}
	\includegraphics[width=0.47\textwidth]{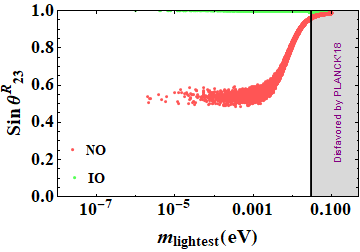}
	\caption{Correlations between parameters using analytical relations followed from $m_{ee}=0, m_{e\mu}=0$.} \label{fig0a}
\end{figure}
\begin{figure}[h]
	\includegraphics[width=0.47\textwidth]{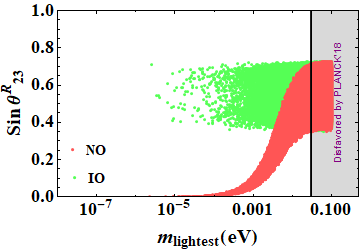}
		\includegraphics[width=0.47\textwidth]{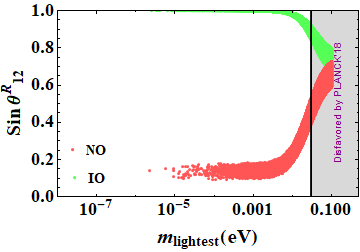}
				\includegraphics[width=0.47\textwidth]{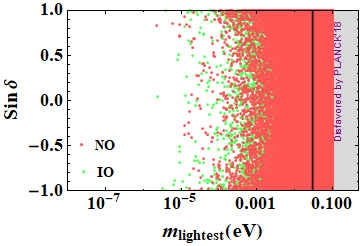}
						\includegraphics[width=0.47\textwidth]{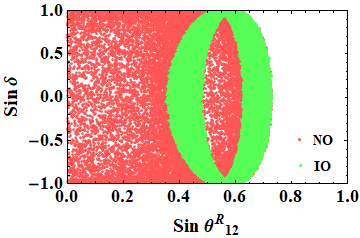}
	\caption{Correlations between parameters using analytical relations followed from $m_{\mu e}=0, m_{\mu \mu}=0$.} \label{fig0b}
\end{figure}

\begin{figure}[h]
	\includegraphics[width=0.47\textwidth]{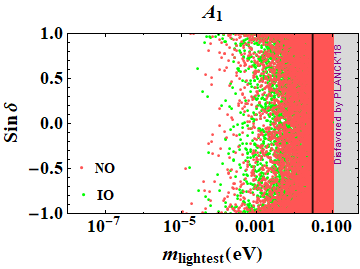}
	\includegraphics[width=0.47\textwidth]{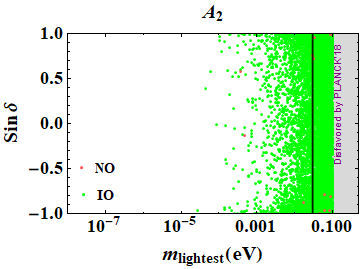}
	\caption{Correlations between parameters for different allowed classes for one-zero texture.} \label{fig1}
\end{figure}
\begin{figure}[h]
	\includegraphics[width=0.47\textwidth]{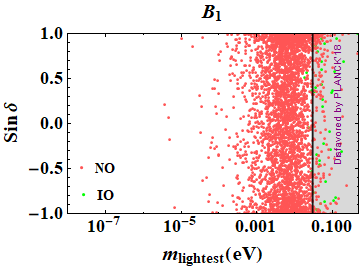}
		\includegraphics[width=0.47\textwidth]{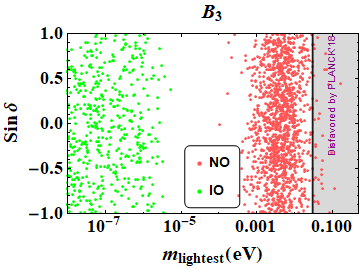}
				\includegraphics[width=0.47\textwidth]{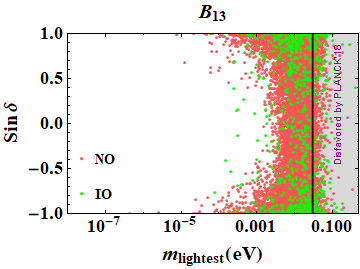}
						\includegraphics[width=0.47\textwidth]{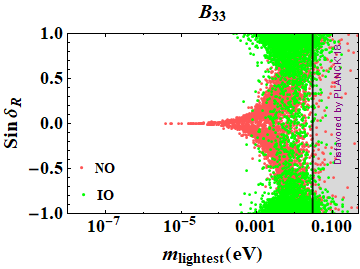}
	\caption{Correlations between parameters for different allowed classes for two-zero texture.} \label{fig2}
\end{figure}

\begin{figure}[h]
	\includegraphics[width=0.47\textwidth]{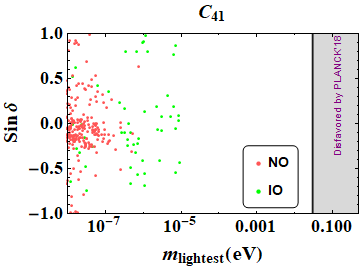}
	\includegraphics[width=0.47\textwidth]{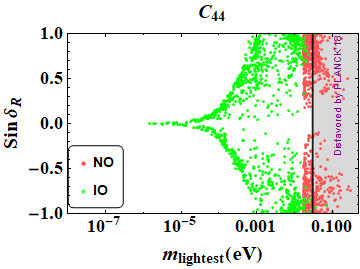}
		\includegraphics[width=0.47\textwidth]{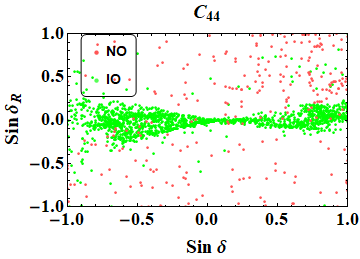}
	\includegraphics[width=0.47\textwidth]{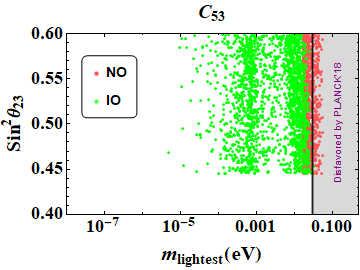}
		\includegraphics[width=0.47\textwidth]{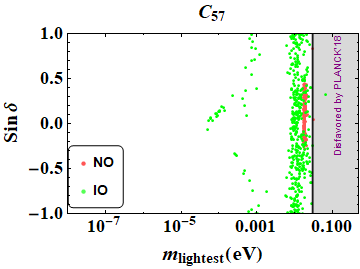}
	\includegraphics[width=0.47\textwidth]{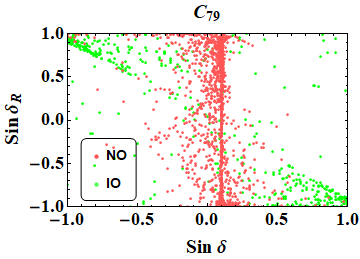}
	\caption{Correlations between parameters for different allowed classes for three-zero texture.} \label{fig3}
\end{figure}

\section{Results and Conclusion}
\label{result}
We start our analysis with the Hermitian light neutrino mass matrix discussed earlier. We first analytically consider a few cases and realise that they lead to unacceptable values of light neutrino parameters. We then move onto numerical calculations for all possible texture zeros in Hermitian case. In this case, the light neutrino mass matrix has only two free parameters namely, the lightest neutrino mass and Dirac CP phase. While texture zero conditions for diagonal entries can be solved for one parameter only, the off diagonal texture zero conditions can be solved for two parameters. We consider both normal and inverted ordering of light neutrino masses, use the $3\sigma$ values of three mixing angles, two mass squared differences to solve the texture zero conditions. We find that all six possible one-zero textures are ruled out by latest neutrino data. This automatically rules out higher textures like two-zero textures. This is not surprising given the fact that almost all Majorana two-zero textures are disallowed by present data \cite{Borgohain:2020now} even though there exists two more CP phases as free parameters in Majorana neutrino mass matrix compared to Dirac Hermitian mass matrix we discuss here.

We now move onto the discussion of non-Hermitian or general structure of light Dirac neutrino mass matrix. Before performing a complete numerical scan over all possible texture zeros in such a case, we first find simple analytical relations which follow from texture zero conditions. The simple analytical relations that follow from texture zero conditions $m_{ee}=0, m_{e\mu}=0$ \eqref{twozero1} lead to interesting correlations between the mixing angles in unitary matrix $V_R$ and the lightest neutrino mass as seen in figure \ref{fig0a}. Similarly, the analytical expressions which follow from texture zero conditions $m_{\mu e}=0, m_{\mu \mu}=0$ \eqref{twozero2} lead to even more interesting correlations due to the presence of leptonic Dirac CP phase $\delta$. They are shown in figure \ref{fig0b}. In our numerical analysis to be discussed below, we find that the results match with our analytical results.

We then numerically check the validity of texture zeros in non-Hermitian or general structure of Dirac neutrino mass matrix. As mentioned earlier, we have many more possible texture zeros in such a scenario. There are $^9C_1=9$ possible one-zero textures, $^9C_2=36$ possible two-zero textures, $^9C_3=84$ possible three-zero textures, $^9C_4=126$ possible four-zero textures and so on. Due to the introduction of another $3\times 3$ unitary matrix, there arises 9 additional free parameters namely, the newly introduced phases $\omega_i, \delta_R \in (-\pi, \pi)$, and mixing angles $\theta^R_{ij} \in (0, \pi/2)$. While solving the texture zero conditions we either vary these new parameters randomly in the entire allowed range or solve for them while demanding the solution to lie in this range, depending on the number of equations and variables. As expected, all one-zero textures are now allowed, as summarised in table \ref{t1}. Similar results for two-zero textures are summarised in table \ref{t2}. Although we expected all of them to be allowed due to the presence of many eleven free parameters and only two constraint equations, yet it is interesting to find that the present data can still rule out texture $B_9$ for inverted mass ordering. Out of 84 three-zero texture possibilities we found only 52 are allowed for both mass ordering while  10 are allowed with only NO, as summarised in table \ref{t22} and table \ref{t222}. On the other hand out of 126 four-zero texture possibilities, only two textures are allowed for both mass ordering while four more are allowed only for NO, as summarised in table \ref{t3}.

Figure \ref{fig1} shows some interesting predictions for leptonic Dirac CP phase and the lightest neutrino mass in case of one-zero texture. While $A_1$ does not discriminate between NO and IO, the texture $A_2$ seems to favour IO more compared to NO, as can be seen from right panel of figure \ref{fig1}. In this figure, we also show the Planck 2018 cosmological bound on $\sum_i \lvert m_i \rvert < 0.12$ eV, by translating it into the corresponding bound on lightest neutrino mass. As one can see, several points predicted by these two particular textures are already disfavoured from cosmology bound. Similar predictions for neutrino parameters in case of two-zero and three-zero textures are shown in figure \ref{fig2} and \ref{fig3} respectively. Sharp distinction between NO and IO is clearly visible once again in two-zero textures as well, for example in $B_1, B_3$ of figure \ref{fig2}. While $B_1$ almost disfavours IO, $B_3$ pushes some of the points for NO to saturate or exceed the upper limit on lightest neutrino mass from cosmology. Similar distinctions are also noticed in three-zero textures shown in figure \ref{fig3}. Apart from predicting the lightest neutrino mass, Dirac CP phase $\delta$, the textures can also predict the values of of some parameters in the mixing matrix $U_{\nu R} $. As an example, we show the predictions for CP phase $\delta_R$ for two-zero texture $B_{33}$ and three-zero textures $C_{44}, C_{51}, C_{79}$ in figure \ref{fig2} and \ref{fig3} respectively. It is to be noted that the correlation between Dirac CP phase and the lightest neutrino mass from our numerical analysis matches with the ones obtained from analytical relations shown in figure \ref{fig0b}.

To conclude, we have made a survey of possible texture zeros in light neutrino mass matrix by considering it to be of Dirac nature and assuming a diagonal charged lepton basis. We first considered the mass matrix to be Hermitian so that it is diagonalisable by a unitary matrix, which is same as the leptonic mixing matrix in diagonal charged lepton basis. Using the latest neutrino global fit $3\sigma$ data we find that none of the one-zero and two-zero textures are allowed which automatically rules out the higher texture zeros as well. We then relax the assumption of Hermitian mass matrix and consider it to be a general complex matrix which can be diagonalised by a bi-unitary transformation. Due to the introduction of additional CP phases and mixing angles, which are not constrained by neutrino data, we get many more possible texture zeros. Apart from classifying the allowed, disallowed textures in this category, we find that neutrino mass matrix with more than four zeros are disallowed by present neutrino data. The allowed textures belonging to one-zero, two-zero, three-zero and four-zero classes also show interesting differences between normal and inverted ordering of light neutrino masses. This result is in sharp contrast with texture zero results of Majorana neutrino scenario where only one two-zero texture and no textures with more than two zeros are allowed \cite{Borgohain:2020now}. It will be interesting to study flavour models of Dirac neutrinos which predict such texture zeros in neutrino mass matrix, while the flavour symmetries predicting more than three texture zeros will be disfavoured by present data, as we find here. Such flavour models often introduce new interactions of right handed neutrinos which can lead to their thermalisation in the early universe leaving interesting signatures at the cosmic microwave background radiation via their contribution to effective relativistic degrees of freedom  \cite{Nanda:2019nqy, Borah:2020boy}. We leave such detailed study of Dirac neutrino textures from flavour symmetric point of view and relevant phenomenology to future works.

\acknowledgments
The authors acknowledge the support from Early Career Research Award from the department of science and technology-science and engineering research board (DST-SERB), Government of India (reference number: ECR/2017/001873)

\appendix
\section{Elements of the light neutrino mass matrix for the Hermitian case}
\label{appen1}
{\small \begin{widetext}
\begin{eqnarray}
m_{ee}=c_{13}^2 \left(c_{12}^2 m_1+m_2 s_{12}^2\right)+m_3 s_{13}^2 \nonumber
\end{eqnarray}
\begin{eqnarray}
m_{e\mu}=c_{13} \left(c_{12} c_{23} \left(m_2-m_1\right) s_{12}+e^{-i \delta } s_{13} s_{23}
   \left(c_{12}^2 \left(-m_1\right)-m_2 s_{12}^2+m_3\right)\right) \nonumber 
\end{eqnarray}
\begin{eqnarray}
m_{e\tau}=c_{13} \left(c_{12} \left(m_1-m_2\right) s_{12} s_{23}+c_{23} e^{-i \delta } s_{13}
   \left(c_{12}^2 \left(-m_1\right)-m_2 s_{12}^2+m_3\right)\right) \nonumber
\end{eqnarray}
\begin{eqnarray}
m_{\mu\mu}=2 c_{12} c_{23} \left(m_1-m_2\right) s_{12} s_{13} s_{23} \cos (\delta )+c_{23}^2
   \left(c_{12}^2 m_2+m_1 s_{12}^2\right)+s_{23}^2 \left(s_{13}^2 \left(c_{12}^2 m_1+m_2
   s_{12}^2\right)+c_{13}^2 m_3\right) \nonumber
\end{eqnarray}
\begin{eqnarray}
m_{\mu\tau}=c_{12} e^{-i \delta } \left(m_1-m_2\right) s_{12} s_{13} \left(c_{23}^2-e^{2 i \delta }
   s_{23}^2\right)+c_{23} c_{12}^2 s_{23} \left(m_1 s_{13}^2-m_2\right)+c_{23} s_{23}
   \left(c_{13}^2 m_3+s_{12}^2 \left(m_2 s_{13}^2-m_1\right)\right) \nonumber
\end{eqnarray}
\begin{eqnarray}
m_{\tau\tau}=2 c_{12} c_{23} \left(m_2-m_1\right) s_{12} s_{13} s_{23} \cos (\delta )+c_{23}^2
   \left(s_{13}^2 \left(c_{12}^2 m_1+m_2 s_{12}^2\right)+c_{13}^2 m_3\right)+s_{23}^2
   \left(c_{12}^2 m_2+m_1 s_{12}^2\right)
 \nonumber
\end{eqnarray}
\end{widetext}}
Also, due to the Hermitian nature, we have $m_{\mu e} = m^*_{e \mu}, m_{\tau e} = m^*_{e \tau}, m_{\tau \mu} = m^*_{\mu \tau}$.
\section{Elements of the light neutrino mass matrix for the non-Hermitian case}
\label{appen2}
Denote the elements of unitary matrices $U_L, V_R$ as
\begin{equation}\label{eq8a}
	U_L=\left[\begin{array}{ccc}
	U_{11} & U_{12} & U_{13} \\
	U_{21} & U_{22} & U_{23} \\
	U_{31} & U_{32} & U_{33}
	\end{array} \right],
	\end{equation}
\begin{equation}\label{eq8a}
	V_R=\left[\begin{array}{ccc}
	V_{11} & V_{12} & V_{13} \\
	V_{21} & V_{22} & V_{23} \\
	V_{31} & V_{32} & V_{33}
	\end{array}\right].
	\end{equation}
The light neutrino mass matrix elements parametrised as
	\begin{equation}\label{eq6}
	M_\nu= U_{\nu L} {M_\nu}^{(\rm diag)} U^{\dagger}_{\nu R}
	\end{equation}
with $U_{\nu L}  = U_L$ and 
\begin{equation}
U_{\nu R} = {\rm diag} \left (e^{i \omega_1}, e^{i \omega_2}, e^{i \omega_3} \right) V_R (\theta^R_{ij}, \delta_R) {\rm diag} \left (e^{i \omega_4}, e^{i \omega_5}, 1 \right).
\end{equation}
The light neutrino mass matrix elements are 
\begin{equation}
m_{ee} = e^{-i(\omega_1+\omega_4+\omega_5)} \bigg [ e^{i\omega_5} U_{11} V^*_{11}m_1+e^{i \omega_4} (U_{12}V^*_{12}m_2+e^{i \omega_5} U_{13} V^*_{13} m_3) \bigg ],
\end{equation}
\begin{equation}
m_{e\mu} = e^{-i(\omega_2+\omega_4+\omega_5)} \bigg [ e^{i\omega_5} U_{11} V^*_{21}m_1+e^{i \omega_4} (U_{12}V^*_{22}m_2+e^{i \omega_5} U_{13} V^*_{23} m_3) \bigg ],
\end{equation}
\begin{equation}
m_{e \tau} = e^{-i(\omega_3+\omega_4+\omega_5)} \bigg [ e^{i\omega_5} U_{11} V^*_{31}m_1+e^{i \omega_4} (U_{12}V^*_{32}m_2+e^{i \omega_5} U_{13} V^*_{33} m_3) \bigg ],
\end{equation}
\begin{equation}
m_{\mu e} = e^{-i(\omega_1+\omega_4+\omega_5)} \bigg [ e^{i\omega_5} U_{21} V^*_{11}m_1+e^{i \omega_4} (U_{22}V^*_{12}m_2+e^{i \omega_5} U_{23} V^*_{13} m_3) \bigg ],
\end{equation}
\begin{equation}
m_{\mu \mu} = e^{-i(\omega_2+\omega_4+\omega_5)} \bigg [ e^{i\omega_5} U_{21} V^*_{21}m_1+e^{i \omega_4} (U_{22}V^*_{22}m_2+e^{i \omega_5} U_{23} V^*_{23} m_3) \bigg ],
\end{equation}
\begin{equation}
m_{\mu \tau} = e^{-i(\omega_3+\omega_4+\omega_5)} \bigg [ e^{i\omega_5} U_{21} V^*_{31}m_1+e^{i \omega_4} (U_{22}V^*_{32}m_2+e^{i \omega_5} U_{23} V^*_{33} m_3) \bigg ],
\end{equation}
\begin{equation}
m_{\tau e} = e^{-i(\omega_1+\omega_4+\omega_5)} \bigg [ e^{i\omega_5} U_{31} V^*_{11}m_1+e^{i \omega_4} (U_{32}V^*_{12}m_2+e^{i \omega_5} U_{33} V^*_{13} m_3) \bigg ],
\end{equation}
\begin{equation}
m_{\tau \mu} = e^{-i(\omega_2+\omega_4+\omega_5)} \bigg [ e^{i\omega_5} U_{31} V^*_{21}m_1+e^{i \omega_4} (U_{32}V^*_{22}m_2+e^{i \omega_5} U_{33} V^*_{23} m_3) \bigg ],
\end{equation}
\begin{equation}
m_{\tau \tau} = e^{-i(\omega_3+\omega_4+\omega_5)} \bigg [ e^{i\omega_5} U_{31} V^*_{31}m_1+e^{i \omega_4} (U_{32}V^*_{32}m_2+e^{i \omega_5} U_{33} V^*_{33} m_3) \bigg ],
\end{equation}
which in terms of mixing angles can be written as
{\small \begin{widetext}
\begin{equation}
m_{ee}=e^{-i \left(\delta +\omega _1+\omega _4+\omega _5\right)} \left(c_{13} \text{cr}_{13}
   e^{i \delta } \left(c_{12} \text{cr}_{12} m_1 e^{i \omega _5}+m_2 s_{12}
   \text{sr}_{12} e^{i \omega _4}\right)+m_3 s_{13} \text{sr}_{13} e^{i \left(\delta
   _R+\omega _4+\omega _5\right)}\right), \nonumber
\end{equation}		
\begin{align}
m_{e\mu}&=c_{13} e^{-i \left(\delta _R+\omega _2+\omega _4+\omega _5\right)} \left(m_2 s_{12} e^{i
   \omega _4} \left(-\text{sr}_{12} \text{sr}_{13} \text{sr}_{23}+\text{cr}_{12}
   \text{cr}_{23} e^{i \delta _R}\right)-c_{12} m_1 e^{i \omega _5} \left(\text{cr}_{12}
   \text{sr}_{13} \text{sr}_{23}+\text{cr}_{23} \text{sr}_{12} e^{i \delta
   _R}\right)\right) \nonumber \\
& +\text{cr}_{13} m_3 s_{13} \text{sr}_{23} e^{-i \left(\delta +\omega
   _2\right)}, \nonumber
\end{align}
\begin{align}
m_{e\tau}&=\text{cr}_{13} \text{cr}_{23} m_3 s_{13} e^{-i \left(\delta +\omega _3\right)}-c_{13}
   c_{12} m_1 e^{-i \left(\delta _R+\omega _3+\omega _4\right)}
   \left(\text{cr}_{12} \text{cr}_{23} \text{sr}_{13}-\text{sr}_{12} \text{sr}_{23} e^{i
   \delta _R}\right) \nonumber \\
&   +m_2 c_{13} s_{12} e^{-i \left(\delta _R+\omega _3+\omega _5\right)}
   \left(\text{cr}_{23} \text{sr}_{12} \text{sr}_{13}+\text{cr}_{12} \text{sr}_{23} e^{i
   \delta _R}\right), \nonumber
\end{align}

\begin{align}
m_{\mu e}&=e^{-i \left(\omega _1+\omega _4+\omega _5\right)} \bigg (s_{23} \left(c_{12}
   \text{cr}_{12} \text{cr}_{13} m_1 s_{13} \left(-e^{i \left(\delta +\omega
   _5\right)}\right)+c_{13} m_3 \text{sr}_{13} e^{i \left(\delta _R+\omega _4+\omega
   _5\right)}-\text{cr}_{13} m_2 s_{12} s_{13} \text{sr}_{12} e^{i \left(\delta +\omega
   _4\right)}\right) \nonumber \\
&  -c_{23} \text{cr}_{13} \left(\text{cr}_{12} m_1 s_{12} e^{i \omega
   _5}-c_{12} m_2 \text{sr}_{12} e^{i \omega _4}\right)\bigg ),   \nonumber
\end{align}
\begin{align}
m_{\mu\mu}&=e^{-i \left(\delta _R+\omega _2+\omega _4+\omega _5\right)} \bigg (m_2 e^{i \omega _4}
   \left(c_{12} c_{23}-e^{i \delta } s_{12} s_{13} s_{23}\right) \left(-\text{sr}_{12}
   \text{sr}_{13} \text{sr}_{23}+\text{cr}_{12} \text{cr}_{23} e^{i \delta_R}\right) \nonumber \\
&   +e^{i \omega _5} \left(c_{13} \text{cr}_{13} m_3 s_{23} \text{sr}_{23} e^{i
   \left(\delta _R+\omega _4\right)}+m_1 \left(c_{23} s_{12}+c_{12} e^{i \delta } s_{13}
   s_{23}\right) \left(\text{cr}_{12} \text{sr}_{13} \text{sr}_{23}+\text{cr}_{23}
   \text{sr}_{12} e^{i \delta _R}\right)\right)\bigg ), \nonumber
\end{align}
\begin{align}
m_{\mu\tau}&=e^{-i \left(\delta _R+\omega _3+\omega _4+\omega _5\right)} \bigg (c_{13} \text{cr}_{13}
   \text{cr}_{23} m_3 s_{23} e^{i \left(\delta _R+\omega _4+\omega
   _5\right)}+\text{sr}_{23} e^{i \delta _R} \bigg (\text{cr}_{12} m_2 \left(-e^{i \omega
   _4}\right) \left(c_{12} c_{23}-e^{i \delta } s_{12} s_{13} s_{23}\right) \nonumber \\
&   -m_1\text{sr}_{12} e^{i \omega _5} \left(c_{23} s_{12}+c_{12} e^{i \delta } s_{13}
   s_{23}\right)\bigg )+\text{cr}_{23} \text{sr}_{13} \bigg (\text{cr}_{12} m_1 e^{i
   \omega _5} \left(c_{23} s_{12}+c_{12} e^{i \delta } s_{13} s_{23}\right) \nonumber \\
   & -m_2
   \text{sr}_{12} e^{i \omega _4} \left(c_{12} c_{23}-e^{i \delta } s_{12} s_{13}
   s_{23}\right)\bigg )\bigg ), \nonumber
\end{align}
\begin{align}
m_{\tau e}&=e^{-i \left(\omega _1+\omega _4+\omega _5\right)} \bigg (\text{cr}_{13} s_{12}
   \left(\text{cr}_{12} m_1 s_{23} e^{i \omega _5}-c_{23} m_2 s_{13} \text{sr}_{12} e^{i
   \left(\delta +\omega _4\right)}\right)-c_{12} \text{cr}_{13} \bigg (c_{23}
   \text{cr}_{12} m_1 s_{13} e^{i \left(\delta +\omega _5\right)} \nonumber \\
   &   +m_2 s_{23}
   \text{sr}_{12} e^{i \omega _4}\bigg )+c_{13} c_{23} m_3 \text{sr}_{13} e^{i
   \left(\delta _R+\omega _4+\omega _5\right)}\bigg ), \nonumber
\end{align}
\begin{align}
m_{\tau\mu}&=e^{-i \left(\delta _R+\omega _2+\omega _4+\omega _5\right)} \bigg (m_1 e^{i \omega _5}
   \left(-s_{12} s_{23}+c_{12} c_{23} e^{i \delta } s_{13}\right) \left(\text{cr}_{12}
   \text{sr}_{13} \text{sr}_{23}+\text{cr}_{23} \text{sr}_{12} e^{i \delta
   _R}\right) \nonumber \\
&   +e^{i \omega _4} \left(c_{13} c_{23} \text{cr}_{13} m_3 \text{sr}_{23} e^{i
   \left(\delta _R+\omega _5\right)}-m_2 \left(c_{12} s_{23}+c_{23} e^{i \delta } s_{12}
   s_{13}\right) \left(-\text{sr}_{12} \text{sr}_{13} \text{sr}_{23}+\text{cr}_{12}
   \text{cr}_{23} e^{i \delta _R}\right)\right)\bigg ), \nonumber
\end{align}
\begin{align}
m_{\tau\tau}& =e^{-i \left(\delta _R+\omega _3+\omega _4+\omega _5\right)} \bigg (c_{13} c_{23}
   \text{cr}_{13} \text{cr}_{23} m_3 e^{i \left(\delta _R+\omega _4+\omega
   _5\right)}+\text{sr}_{23} e^{i \delta _R} \bigg (\text{cr}_{12} m_2 e^{i \omega _4}
   \left(c_{12} s_{23}+c_{23} e^{i \delta } s_{12} s_{13}\right) \nonumber \\
&   -m_1 \text{sr}_{12} e^{i
   \omega _5} \left(-s_{12} s_{23}+c_{12} c_{23} e^{i \delta }
   s_{13}\right)\bigg )+\text{cr}_{23} \text{sr}_{13} \bigg (\text{cr}_{12} m_1 e^{i
   \omega _5} \left(-s_{12} s_{23}+c_{12} c_{23} e^{i \delta } s_{13}\right) \nonumber \\
&   +m_2
   \text{sr}_{12} e^{i \omega _4} \left(c_{12} s_{23}+c_{23} e^{i \delta } s_{12}
   s_{13}\right)\bigg )\bigg ). \nonumber
\end{align}
\end{widetext}}

\bibliographystyle{JHEP}
\bibliography{TODLMA} 

\end{document}